\documentclass[twocolumn,letterpaper,aps,prd,longbibliography,superscriptaddress,nofootinbib,floatfix]{revtex4-2}

\usepackage{graphicx}	
\usepackage{xcolor}
\usepackage{xspace}	

\newcommand{\pt}{\mbox{$p_T$}\xspace}

\newcommand{\sqs}{\mbox{$\sqrt{s}$}\xspace}

\newcommand{\pp}{\mbox{$p$$+$$p$}\xspace}

\newcommand{\antikt}{\mbox{anti-}$k_{t}$\xspace}

\begin{document}

\title{Measurement of inclusive jet cross section and substructure in 
$p$$+$$p$ collisions at $\sqrt{s}=200$ GeV}

\newcommand{\abilene}{Abilene Christian University, Abilene, Texas 79699, USA}
\newcommand{\augie}{Department of Physics, Augustana University, Sioux Falls, South Dakota 57197, USA}
\newcommand{\banaras}{Department of Physics, Banaras Hindu University, Varanasi 221005, India}
\newcommand{\barc}{Bhabha Atomic Research Centre, Bombay 400 085, India}
\newcommand{\baruch}{Baruch College, City University of New York, New York, New York, 10010 USA}
\newcommand{\bnlcoll}{Collider-Accelerator Department, Brookhaven National Laboratory, Upton, New York 11973-5000, USA}
\newcommand{\bnlphys}{Physics Department, Brookhaven National Laboratory, Upton, New York 11973-5000, USA}
\newcommand{\caucr}{University of California-Riverside, Riverside, California 92521, USA}
\newcommand{\charlesczech}{Charles University, Faculty of Mathematics and Physics, 180 00 Troja, Prague, Czech Republic}
\newcommand{\ciae}{Science and Technology on Nuclear Data Laboratory, China Institute of Atomic Energy, Beijing 102413, People's Republic of China}
\newcommand{\cns}{Center for Nuclear Study, Graduate School of Science, University of Tokyo, 7-3-1 Hongo, Bunkyo, Tokyo 113-0033, Japan}
\newcommand{\colorado}{University of Colorado, Boulder, Colorado 80309, USA}
\newcommand{\columbia}{Columbia University, New York, New York 10027 and Nevis Laboratories, Irvington, New York 10533, USA}
\newcommand{\czechtech}{Czech Technical University, Zikova 4, 166 36 Prague 6, Czech Republic}
\newcommand{\debrecen}{Debrecen University, H-4010 Debrecen, Egyetem t{\'e}r 1, Hungary}
\newcommand{\elte}{ELTE, E{\"o}tv{\"o}s Lor{\'a}nd University, H-1117 Budapest, P{\'a}zm{\'a}ny P.~s.~1/A, Hungary}
\newcommand{\ewha}{Ewha Womans University, Seoul 120-750, Korea}
\newcommand{\fsu}{Florida State University, Tallahassee, Florida 32306, USA}
\newcommand{\gsu}{Georgia State University, Atlanta, Georgia 30303, USA}
\newcommand{\hanyang}{Hanyang University, Seoul 133-792, Korea}
\newcommand{\hiroshima}{Physics Program and International Institute for Sustainability with Knotted Chiral Meta Matter (WPI-SKCM$^2$), Hiroshima University, Higashi-Hiroshima, Hiroshima 739-8526, Japan}
\newcommand{\howard}{Department of Physics and Astronomy, Howard University, Washington, DC 20059, USA}
\newcommand{\hunrenatomki}{HUN-REN ATOMKI, H-4026 Debrecen, Bem t{\'e}r 18/c, Hungary}
\newcommand{\ihepprot}{IHEP Protvino, State Research Center of Russian Federation, Institute for High Energy Physics, Protvino, 142281, Russia}
\newcommand{\illuiuc}{University of Illinois at Urbana-Champaign, Urbana, Illinois 61801, USA}
\newcommand{\inrras}{Institute for Nuclear Research of the Russian Academy of Sciences, prospekt 60-letiya Oktyabrya 7a, Moscow 117312, Russia}
\newcommand{\instpasczech}{Institute of Physics, Academy of Sciences of the Czech Republic, Na Slovance 2, 182 21 Prague 8, Czech Republic}
\newcommand{\isu}{Iowa State University, Ames, Iowa 50011, USA}
\newcommand{\jaea}{Advanced Science Research Center, Japan Atomic Energy Agency, 2-4 Shirakata Shirane, Tokai-mura, Naka-gun, Ibaraki-ken 319-1195, Japan}
\newcommand{\jeonbuk}{Jeonbuk National University, Jeonju, 54896, Korea}
\newcommand{\jyvaskyla}{Helsinki Institute of Physics and University of Jyv{\"a}skyl{\"a}, P.O.Box 35, FI-40014 Jyv{\"a}skyl{\"a}, Finland}
\newcommand{\kek}{KEK, High Energy Accelerator Research Organization, Tsukuba, Ibaraki 305-0801, Japan}
\newcommand{\korea}{Korea University, Seoul 02841, Korea}
\newcommand{\kurchatov}{National Research Center ``Kurchatov Institute", Moscow, 123098 Russia}
\newcommand{\kyoto}{Kyoto University, Kyoto 606-8502, Japan}
\newcommand{\labllr}{Laboratoire Leprince-Ringuet, Ecole Polytechnique, CNRS-IN2P3, Route de Saclay, F-91128, Palaiseau, France}
\newcommand{\lahorelums}{Physics Department, Lahore University of Management Sciences, Lahore 54792, Pakistan}
\newcommand{\lawllnl}{Lawrence Livermore National Laboratory, Livermore, California 94550, USA}
\newcommand{\losalamos}{Los Alamos National Laboratory, Los Alamos, New Mexico 87545, USA}
\newcommand{\lund}{Department of Physics, Lund University, Box 118, SE-221 00 Lund, Sweden}
\newcommand{\lyon}{IPNL, CNRS/IN2P3, Univ Lyon, Universit{\'e} Lyon 1, F-69622, Villeurbanne, France}
\newcommand{\maryland}{University of Maryland, College Park, Maryland 20742, USA}
\newcommand{\mass}{Department of Physics, University of Massachusetts, Amherst, Massachusetts 01003-9337, USA}
\newcommand{\mate}{MATE, Laboratory of Femtoscopy, K\'aroly R\'obert Campus, H-3200 Gy\"ongy\"os, M\'atrai\'ut 36, Hungary}
\newcommand{\michigan}{Department of Physics, University of Michigan, Ann Arbor, Michigan 48109-1040, USA}
\newcommand{\miss}{Mississippi State University, Mississippi State, Mississippi 39762, USA}
\newcommand{\muhlenberg}{Muhlenberg College, Allentown, Pennsylvania 18104-5586, USA}
\newcommand{\myongji}{Myongji University, Yongin, Kyonggido 449-728, Korea}
\newcommand{\nagasaki}{Nagasaki Institute of Applied Science, Nagasaki-shi, Nagasaki 851-0193, Japan}
\newcommand{\nara}{Nara Women's University, Kita-uoya Nishi-machi Nara 630-8506, Japan}
\newcommand{\natmephi}{National Research Nuclear University, MEPhI, Moscow Engineering Physics Institute, Moscow, 115409, Russia}
\newcommand{\newmex}{University of New Mexico, Albuquerque, New Mexico 87131, USA}
\newcommand{\nmsu}{New Mexico State University, Las Cruces, New Mexico 88003, USA}
\newcommand{\northcg}{Physics and Astronomy Department, University of North Carolina at Greensboro, Greensboro, North Carolina 27412, USA}
\newcommand{\ohio}{Department of Physics and Astronomy, Ohio University, Athens, Ohio 45701, USA}
\newcommand{\ornl}{Oak Ridge National Laboratory, Oak Ridge, Tennessee 37831, USA}
\newcommand{\orsay}{IPN-Orsay, Univ.~Paris-Sud, CNRS/IN2P3, Universit\'e Paris-Saclay, BP1, F-91406, Orsay, France}
\newcommand{\peking}{Peking University, Beijing 100871, People's Republic of China}
\newcommand{\pnpi}{PNPI, Petersburg Nuclear Physics Institute, Gatchina, Leningrad region, 188300, Russia}
\newcommand{\pusan}{Pusan National University, Pusan 46241, Korea}
\newcommand{\riken}{RIKEN Nishina Center for Accelerator-Based Science, Wako, Saitama 351-0198, Japan}
\newcommand{\rikjrbrc}{RIKEN BNL Research Center, Brookhaven National Laboratory, Upton, New York 11973-5000, USA}
\newcommand{\rikkyo}{Physics Department, Rikkyo University, 3-34-1 Nishi-Ikebukuro, Toshima, Tokyo 171-8501, Japan}
\newcommand{\saispbstu}{Saint Petersburg State Polytechnic University, St.~Petersburg, 195251 Russia}
\newcommand{\seoulnat}{Department of Physics and Astronomy, Seoul National University, Seoul 151-742, Korea}
\newcommand{\stonybrkc}{Chemistry Department, Stony Brook University, SUNY, Stony Brook, New York 11794-3400, USA}
\newcommand{\stonycrkp}{Department of Physics and Astronomy, Stony Brook University, SUNY, Stony Brook, New York 11794-3800, USA}
\newcommand{\sungskku}{Sungkyunkwan University, Suwon, 440-746, Korea}
\newcommand{\tenn}{University of Tennessee, Knoxville, Tennessee 37996, USA}
\newcommand{\titech}{Department of Physics, Tokyo Institute of Technology, Oh-okayama, Meguro, Tokyo 152-8551, Japan}
\newcommand{\tsukuba}{Tomonaga Center for the History of the Universe, University of Tsukuba, Tsukuba, Ibaraki 305, Japan}
\newcommand{\usmma}{United States Merchant Marine Academy, Kings Point, New York 11024, USA}
\newcommand{\vandy}{Vanderbilt University, Nashville, Tennessee 37235, USA}
\newcommand{\weizmann}{Weizmann Institute, Rehovot 76100, Israel}
\newcommand{\wigner}{Institute for Particle and Nuclear Physics, HUN-REN Wigner Research Centre for Physics, (HUN-REN Wigner RCP, RMI) H-1525 Budapest 114, POBox 49, Budapest, Hungary}
\newcommand{\yonsei}{Yonsei University, IPAP, Seoul 120-749, Korea}
\newcommand{\zagreb}{Department of Physics, Faculty of Science, University of Zagreb, Bijeni\v{c}ka c.~32 HR-10002 Zagreb, Croatia}
\newcommand{\zambia}{Department of Physics, School of Natural Sciences, University of Zambia, Great East Road Campus, Box 32379, Lusaka, Zambia}
\affiliation{\abilene}
\affiliation{\augie}
\affiliation{\banaras}
\affiliation{\barc}
\affiliation{\baruch}
\affiliation{\bnlcoll}
\affiliation{\bnlphys}
\affiliation{\caucr}
\affiliation{\charlesczech}
\affiliation{\ciae}
\affiliation{\cns}
\affiliation{\colorado}
\affiliation{\columbia}
\affiliation{\czechtech}
\affiliation{\debrecen}
\affiliation{\elte}
\affiliation{\ewha}
\affiliation{\fsu}
\affiliation{\gsu}
\affiliation{\hanyang}
\affiliation{\hiroshima}
\affiliation{\howard}
\affiliation{\hunrenatomki}
\affiliation{\ihepprot}
\affiliation{\illuiuc}
\affiliation{\inrras}
\affiliation{\instpasczech}
\affiliation{\isu}
\affiliation{\jaea}
\affiliation{\jeonbuk}
\affiliation{\jyvaskyla}
\affiliation{\kek}
\affiliation{\korea}
\affiliation{\kurchatov}
\affiliation{\kyoto}
\affiliation{\labllr}
\affiliation{\lahorelums}
\affiliation{\lawllnl}
\affiliation{\losalamos}
\affiliation{\lund}
\affiliation{\lyon}
\affiliation{\maryland}
\affiliation{\mass}
\affiliation{\mate}
\affiliation{\michigan}
\affiliation{\miss}
\affiliation{\muhlenberg}
\affiliation{\myongji}
\affiliation{\nagasaki}
\affiliation{\nara}
\affiliation{\natmephi}
\affiliation{\newmex}
\affiliation{\nmsu}
\affiliation{\northcg}
\affiliation{\ohio}
\affiliation{\ornl}
\affiliation{\orsay}
\affiliation{\peking}
\affiliation{\pnpi}
\affiliation{\pusan}
\affiliation{\riken}
\affiliation{\rikjrbrc}
\affiliation{\rikkyo}
\affiliation{\saispbstu}
\affiliation{\seoulnat}
\affiliation{\stonybrkc}
\affiliation{\stonycrkp}
\affiliation{\sungskku}
\affiliation{\tenn}
\affiliation{\titech}
\affiliation{\tsukuba}
\affiliation{\usmma}
\affiliation{\vandy}
\affiliation{\weizmann}
\affiliation{\wigner}
\affiliation{\yonsei}
\affiliation{\zagreb}
\affiliation{\zambia}
\author{N.J.~Abdulameer} \affiliation{\debrecen} \affiliation{\hunrenatomki}
\author{U.~Acharya} \affiliation{\gsu}
\author{C.~Aidala} \affiliation{\losalamos} \affiliation{\michigan} 
\author{N.N.~Ajitanand} \altaffiliation{Deceased} \affiliation{\stonybrkc} 
\author{Y.~Akiba} \email[PHENIX Spokesperson: ]{akiba@rcf.rhic.bnl.gov} \affiliation{\riken} \affiliation{\rikjrbrc}
\author{R.~Akimoto} \affiliation{\cns} 
\author{J.~Alexander} \affiliation{\stonybrkc} 
\author{M.~Alfred} \affiliation{\howard} 
\author{V.~Andrieux} \affiliation{\michigan} 
\author{S.~Antsupov} \affiliation{\saispbstu}
% ...
\collaboration{PHENIX Collaboration}  \noaffiliation

\date{\today}


\begin{abstract}


The jet cross section and jet-substructure observables in $p$$+$$p$ 
collisions at $\sqrt{s}=200$ GeV were measured by the PHENIX 
Collaboration at the Relativistic Heavy Ion Collider (RHIC). Jets are 
reconstructed from charged-particle tracks and 
electromagnetic-calorimeter clusters using the anti-$k_{t}$ algorithm 
with a jet radius $R=0.3$ for jets with transverse momentum within 
$8.0<p_T<40.0$ GeV/$c$ and pseudorapidity $|\eta|<0.15$.  Measurements 
include the jet cross section, as well as distributions of 
SoftDrop-groomed momentum fraction ($z_g$), charged-particle transverse 
momentum with respect to jet axis ($j_T$), and radial distributions of 
charged particles within jets ($r$).  Also measured was the 
distribution of $\xi=-ln(z)$, where $z$ is the fraction of the jet 
momentum carried by the charged particle. The measurements are compared 
to theoretical next-to and next-to-next-to-leading-order calculations, 
the {\sc pythia} and {\sc Herwig} event generators, and to other 
existing experimental results. Indicated from these measurements is a 
lower particle multiplicity in jets at RHIC energies when compared to 
models.  Also noted are implications for future jet measurements with 
sPHENIX at RHIC as well as at the future Electron-Ion Collider.

\end{abstract}

\maketitle

\section{Introduction}

Jets, the collimated sprays of particles originating from hard parton 
scatterings, were initially conceptualized as a probe of quantum 
chromodynamics (QCD)~\cite{Sterman:1977wj}.  Perturbative QCD (pQCD) is 
broadly in good agreement with measurements of jets produced in 
high-energy collisions, particularly at high momentum and large 
radii~\cite{Kogler:2018hem}.  At lower momenta~\cite{ALICE:2019wqv} and 
small radii~\cite{ALICE:2013yva}, it is necessary to include a good 
description of the nonperturbative contributions to jet production, 
including hadronization.  Jet spectra at low momenta are also sensitive 
to the underlying event and effects such as color 
reconnections~\cite{CDFcolcoh,D0colcoh,ALICE:2019wqv}.  Measurements at 
lower momenta are important to test models for these nonperturbative 
components and their effect on jet production.

The use of jets has been expanded to include measurements of jet 
substructure, with a wide variety of observables sensitive to 
distribution of energy within the jet~\cite{Kogler:2018hem}.  These 
observables are sensitive to final-state radiation patterns in QCD.  At 
the Large Hadron Collider (LHC), where studies are dominated by high 
energy jets, pQCD and models generally reproduce data within 
$\approx$20\% for most substructure 
measurements~\cite{ALICE:2022vsz,ALICE:2022hyz,ALICE:2020pga,ALICE:2019qyj,ALICE:2019wqv,CMS:2022drg,CMS:2022drg}.  
Calculations of some observables at RHIC are likewise within 
$\approx$20\% of the data~\cite{STAR:2020ejj}. However, in lower-energy 
collisions, experimental and theoretical uncertainties have generally 
been large~\cite{STAR:2006opb} and there are some measurements where 
theoretical calculations barely agree with the data within large 
uncertainties~\cite{STAR:2021lvw}.  Monte-Carlo generators and pQCD 
calculations are often used to predict the behavior of jets for 
proposed detectors, to determine corrections to measurements, and as a 
baseline for systems where jets may be modified, such as high-energy 
heavy ion collisions.  Simultaneous comparisons between models and data 
for both cross sections and substructure can place substantial 
constraints on Monte-Carlo models.

At the future Electron-Ion Collider (EIC)~\cite{Willeke_2021}, jets 
will also serve as a tool to study the momentum space structure of 
hadrons as well as parton energy loss in cold nuclear 
matter~\cite{Abdul_Khalek_2022}.  Because the EIC will operate at a 
relatively low center-of-mass energy, closer to RHIC than the LHC, it 
is important to have measurements in \pp collisions at comparable 
energies to test universality and factorization breaking 
effects~\cite{TMDfactorization}.

Here are presented measurements of the jet cross section and several 
jet substructure measurements in $p$$+$$p$ collisions at $\sqrt{s}=200$ 
GeV.  The technique is first summarized, including details of the 
unfolding and the simulations used for detector corrections.  The 
results are then presented and compared to pQCD calculations and 
output from the {\sc pythia} and {\sc Herwig} event generators, and 
finally, the implications of these results are discussed.

\section{Jet reconstruction and unfolding}

\subsection{PHENIX detector and data set}

Combined $p$$+$$p$ data sets collected during 2012 and 2015 
were used in this analysis. The 2012 dataset sampled an integrated 
luminosity of 1.55~$pb^{-1}$ using an electromagnetic-calorimeter 
trigger, while the 2015 dataset sampled 13.5~$pb^{-1}$ using a 
similar trigger with a higher-energy threshold.

Jets were measured in the PHENIX central arms~\cite{Adcox:2003zm}. Each 
arm covers a pseudorapidity range of $|\eta|<0.35$ and an azimuthal 
range of $\pi/2$. Charged-particle tracks were measured by a set of 
multi-wire proportional chambers, including an inner drift chamber and 
multiple outer pad chambers.  Energy deposits from neutral particles 
are measured by the finely segmented electromagnetic calorimeter 
(EMCal), consisting of lead-scintillator modules in the west arm, and 
lead-scintillator and lead-glass \v{C}erenkov modules in the east arm. 
The modules have a resolution determined by beam 
tests~\cite{Gabor:1998,phenixemcal} to be 
$\delta E/E = 8.1\% / \sqrt{E} \oplus 2.1\%$ and $5.9\% / \sqrt{E} 
\oplus 0.8\%$, respectively, where $E$ is in GeV, and were calibrated 
through the reconstruction of neutral pion decays. The calorimeter 
further provides a trigger signal initiated by the presence of at least 
1.6 GeV (2012) or 2.1 GeV (2015) of energy deposited in one of the groups of 
overlapping $4 \times 4$ towers in the lead-glass or lead-scintillator 
modules, respectively. To reduce the inefficiencies introduced by dead 
areas in the outer pad chambers, a confirming hit with an energy 
greater than a MIP is required in the EMCal if a drift chamber track 
does not have a confirming hit in the outer pad chamber. In addition to 
the spectrometer arms, a pair of beam-beam counter (BBC) detectors situated 
along the beam line at $3.0<|\eta|<3.9$ provide the minimum-bias 
(MB) trigger signal and reconstruct the $z$ position of the primary 
vertex. The BBCs measure charged particles and are 
used to determine the collision time and vertex position along the beam 
axis.

\subsection{Jet reconstruction}

Jets were reconstructed using the \antikt 
algorithm~\cite{Cacciari:2008, Cacciari:2012} with radius parameter $R 
= 0.3$ from electromagnetic clusters in the EMCal~\cite{phenix2002} and 
charged-particle tracks (in the drift and pad 
chambers)~\cite{phenixtracking1} each with a minimum \pt of 0.5 
GeV/$c$. The \antikt algorithm, the de facto standard for hadronic 
collisions, was chosen because it clusters outward from the hard core 
of jets, thus reducing the sensitivity to detector edges. A set of 
criteria designed to select charged particles with a well-measured 
momentum, and reject conversions and ghost tracks were applied to 
candidate reconstructed tracks. Clusters consistent with arising from 
the same particle as a reconstructed track were rejected to avoid 
double counting the jet constituent energy. To eliminate both beam and 
detector backgrounds, jets were required to have at least three 
constituents, have a charged fraction of momentum between 0.3 and 0.7, 
be within $|\eta_{\rm jet}|<0.15$, and be reconstructed in the same 
PHENIX detector arm that provided the trigger signal. Only events 
passing the offline-event vertex cut $|z_{\rm vertex}|<10$~cm were 
accepted. Jets were required to be fully contained within the 
$\eta,\phi$ acceptance of the PHENIX arm, where the $\eta$ acceptance 
takes into account the longitudinal vertex location. The reconstructed 
jets average between 45\%--55\% of the true jet energy, with the 
average fraction increasing slowly with jet \pt. A jet at a 
reconstructed \pt of 10~GeV/$c$ has a mean of 4.5 track and cluster 
constituents. This rises to a mean of 6 track and cluster constituents 
at a reconstructed \pt of 20~GeV/$c$.

In addition to the jet cross section, the following substructure 
properties were also measured:

\begin{itemize}

\item distributions of SoftDrop~\cite{Dasgupta:2013,Larkoski:2014} 
groomed momentum fraction ($z_g$),

\item charged-particle transverse momentum with respect to the jet axis 
($j_T$),

\item radial distributions of charged particles within the jet 
($r = \sqrt{\Delta\phi^2 + \Delta\eta^2}$), where $\Delta\phi$ and 
$\Delta\eta$ are the distances from a charged particle to the jet axis in 
azimuthal angle and pseudorapidity respectively, 

\item distributions of $\xi=-ln(z)$, where $z$ is the fraction of the 
jet momentum carried by the charged particle.

\end{itemize}

\subsection{Unfolding} 

The reconstructed jet distributions were corrected for the detector 
response by Bayesian unfolding~\cite{Bayes} using the ``RooUnfold" 
framework~\cite{RooUnfold}. The response matrix for unfolding was 
obtained using a simulation of $p$$+$$p$ collision events with the 
PHENIX detector response simulated by {\sc geant3}~\cite{Brun:1119728}. 
In the first step, {\sc pythia}6~\cite{Pythia} Tune A was used with QCD 
hard-scattering processes selected, along with an additional Gaussian 
partonic $k_t$ smearing with a width of 3~GeV as this combination 
better reproduces two-particle correlations previously measured in 
PHENIX~\cite{Adler2006}. To sample the full jet cross section as a 
function of jet \pt with adequate statistics, the {\sc pythia}6 events 
were generated in fixed ranges of partonic \pt.  The full event sample 
was obtained by recombination using the cross section reported by {\sc 
pythia}6. The PHENIX simulations included run-specific detector 
configurations, including tracking and EMCal tower inefficiency and 
trigger efficiency maps. The simulation response was matched to the 
reconstructed data, and the same kinematic and selection cuts used in 
the data were applied to the reconstructed results in the simulation. A 
total of $\approx$832,000 {\sc pythia}6 events per run configuration 
were processed through the full {\sc geant}-based simulation of the 
PHENIX detector.

In processing the simulated events to generate the response matrix, 
jet finding is performed with both the truth simulation input and the 
reconstructed output of the simulations. Truth jets are determined 
directly from the MC-generator output, excluding neutrinos.  The same 
binning was used for both the truth input and reconstructed output, 
both in jet $p_T$ and in the substructure distribution variables. 
When constructing a response matrix three possibilities need to be 
considered:

\begin{enumerate}

\item A matched reconstructed and truth jet pair is found. This is used 
to define the mapping between the reconstructed and truth \pt and 
substructure quantities.

\item A corresponding reconstructed jet is not found for a given truth 
jet. This jet may not have been found due to detector inefficiencies or 
acceptance limitations, analysis cuts, or failed to satisfy a trigger 
condition. This inefficiency must be accounted for when reconstructing 
the truth \pt spectrum and substructure distributions.

\item A reconstructed jet is found that does not match a truth jet in 
the simulated data. This is a fake jet and typically represents a 
contribution from the underlying event. This contribution is more 
important at low jet \pt, and the contribution of these jets are 
subtracted from the measured jet \pt spectrum and substructure 
distributions as part of the unfolding process.
    
\end{enumerate}

\begin{figure}[thb]
\includegraphics[width=1.0\linewidth]{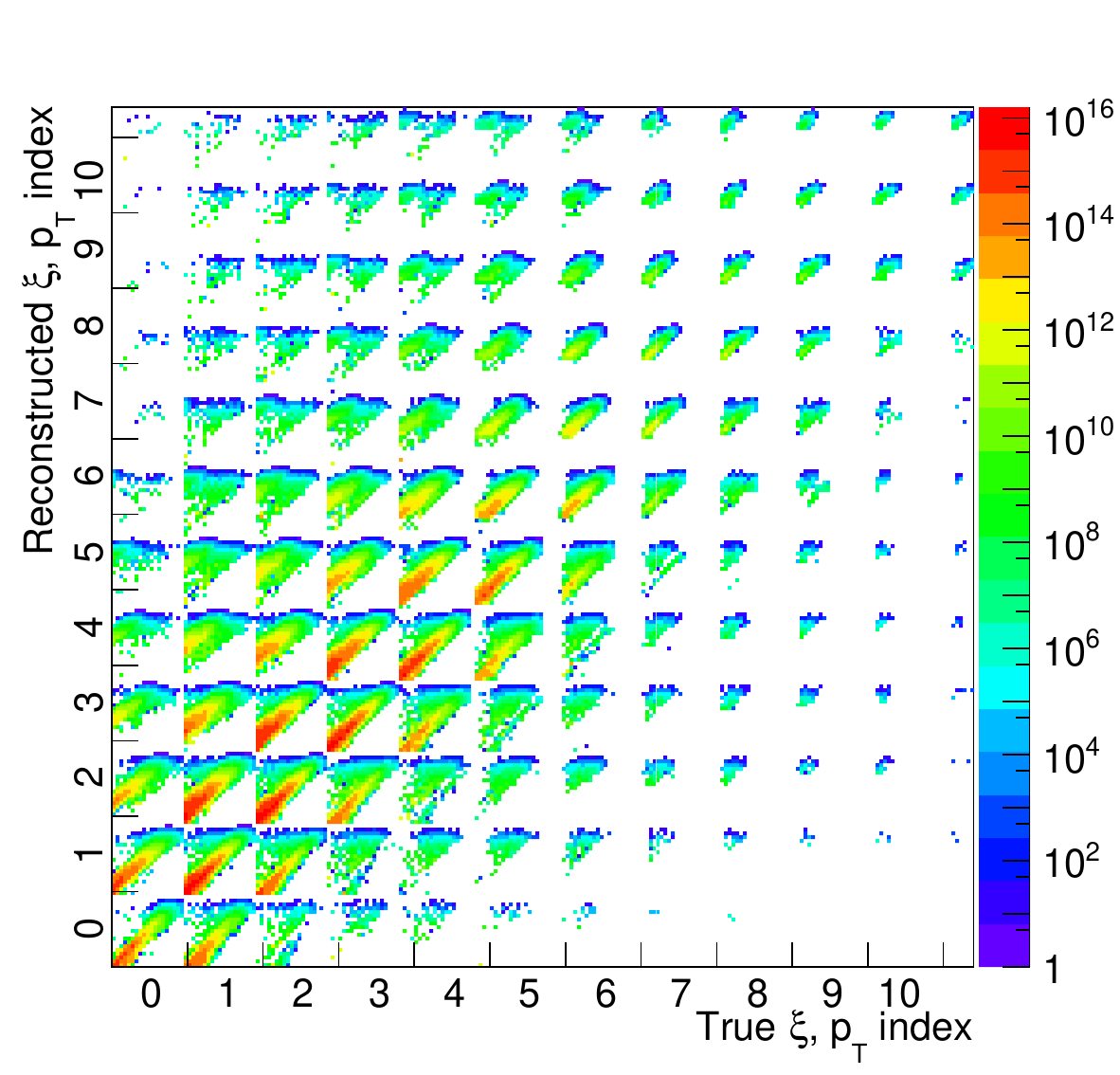}
\caption{\label{fig:unfmat} 
An example of the unfolding matrix for a two-dimensional unfolding of the 
$\xi$ distribution. The matrix is viewed as a set of one-dimensional 
unfolding matrices in jet \pt, shown as a two-dimensional plot in 
reconstructed vs. true jet \pt, as a function of the bin in the 
variable $\xi$.}
\end{figure}

Response matrices are defined for the cross section, which are used for 
one-dimensional unfolding in jet \pt.  For the substructure 
distributions, two dimensional unfolding is performed in both jet \pt 
and the substructure variable. This is a commonly used method which 
takes advantage of a relatively large statistical sample of simulated data to 
extract information from measurements that have a substantial smearing 
in the reconstructed quantities. An example of the unfolding matrix for 
the two-dimensional unfolding for the $\xi$ substructure variable is 
shown in Fig.~\ref{fig:unfmat}.

To test the unfolding procedure a closure test was performed. Two 
statistically independent samples of {\sc pythia}6 simulated events were used 
to determine the response matrices and provide a sample that was 
treated as pseudo-data. The pseudo-data were unfolded following the 
exact same procedures as for data, and the results compared to the 
{\sc pythia}6 truth distributions. The results of the closure test indicate 
that the unfolding method can reproduce the input distribution with a 
high degree of fidelity, and there are no errors in the procedure that 
cause a deviation between the input and unfolded distributions. Tests 
with Bayesian and Singular Value Decomposition (SVD) unfolding using 
{\sc pythia}6 pseudo-data and reweighting the cross section to 
next-to-leading-order (NLO) predictions showed that the unfolding 
converged after two iterations. In what follows Bayesian 
unfolding is used with two iterations, and the difference between the 
second and third iteration is used as a systematic uncertainty.

An examination of the unfolding using the initial {\sc pythia}6 sample 
showed that the integral of the $\xi = -ln(z)$, $j_T$ and $r$ distributions, 
which is the average number of charged particles in the jet, were consistent 
with each other but lower than the {\sc pythia}6 input by approximately one 
charged particle per jet, with a weak dependence on jet \pt.  Similar 
results were obtained using the 
{\sc pythia}8~\cite{bierlich2022comprehensive} event generator in its 
standard configuration with the Monash tune~\cite{Skands:2014pea}. A similar 
discrepancy between measurements and {\sc pythia}6 was observed at RHIC in 
inclusive $\pi^{\pm}$ yields~\cite{Adam2019}. This difference leads to a 
bias in the unfolded substructure distributions, as well as a systematic 
uncertainty in the determination of the efficiencies used to correct the jet 
\pt cross section. As described below, to mitigate this bias the choice is 
made to modify the {\sc pythia}6 events used to generate the unfolding 
matrices in an iterative fashion until the {\sc pythia}6 events better 
matched the unfolded substructure distributions.

\begin{figure}[thb]
\includegraphics[width=1.0\linewidth]{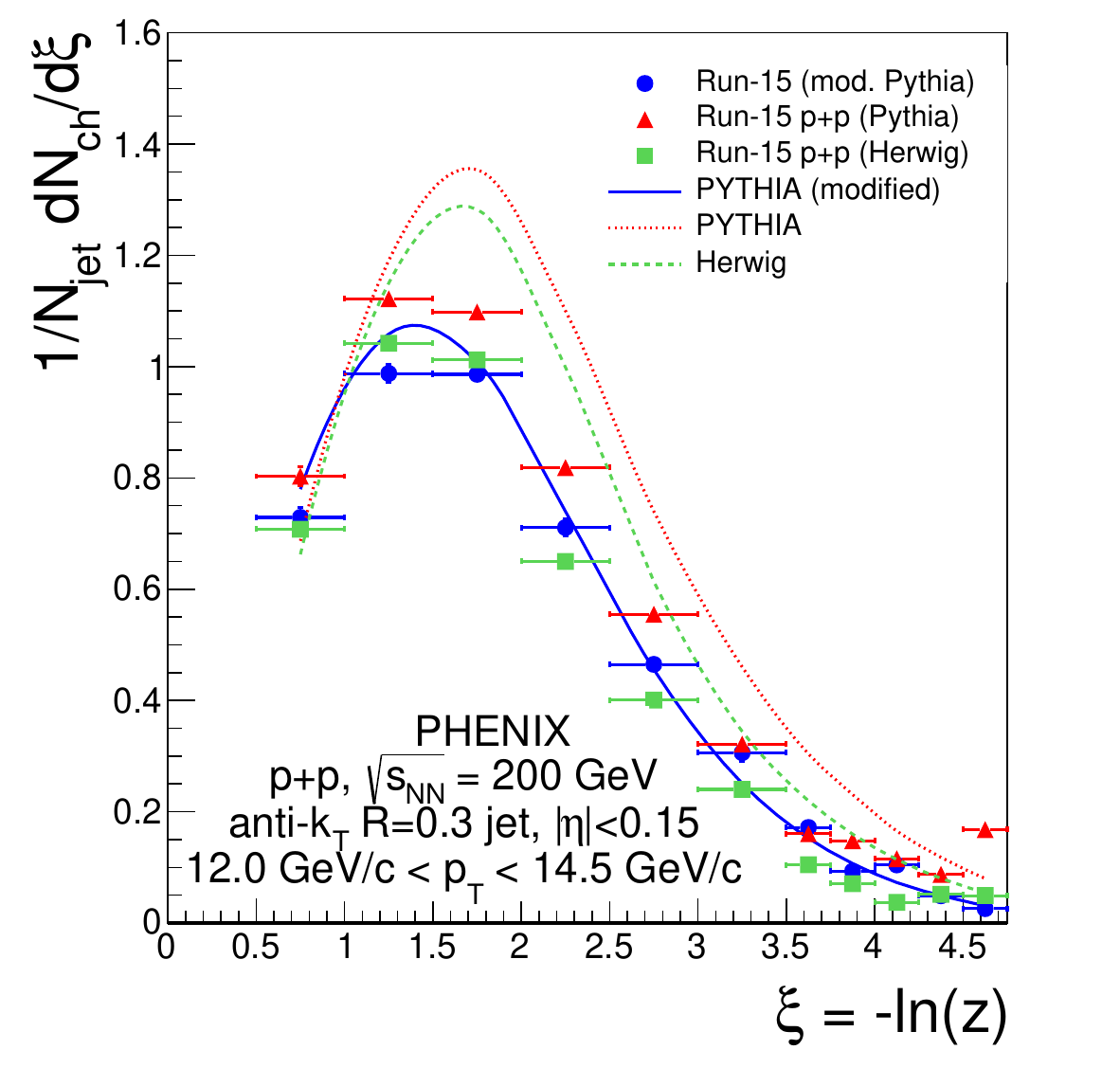}
\caption{\label{fig:pythia68} 
Unfolded $\xi$ distributions (data points) obtained using {\sc Herwig}, 
{\sc pythia} and modified {\sc pythia}6 compared to the 2015 dataset 
results unfolded using the different MC models (dashed lines). Both 
{\sc pythia}6 and {\sc pythia}8 produce compatible results and are 
shown by one curve in the figure. The standard versions of 
{\sc pythia}6 and 8 overestimate the number of jet constituents and 
systematically pull the unfolded distributions to higher values, as 
does the {\sc Herwig} model. The modified version of {\sc pythia}6, 
described in the text, shows a good agreement between the unfolded and 
model distributions and is used for subsequent results in this paper.
}
\end{figure}

As an example, unfolded $\xi$ distributions obtained using different 
versions of {\sc pythia} and {\sc Herwig} are shown in 
Fig.~\ref{fig:pythia68}. The {\sc Herwig} and unmodified {\sc pythia} 
references tend to pull the unfolded distributions systematically high, 
while iteratively adjusting the MC input to be closer to the data 
allows the input distribution to converge to the unfolded data.

Upon examining the difference between the initial unfolding with 
{\sc pythia6} for the charged-particle substructure distributions, it 
was noted that the difference in $r$ between the unfolded data and the 
{\sc pythia6} model was predominantly at large distances from the jet 
axis, which is correlated with a deficit in the unfolded data at large 
$j_T$.  This indicates that a simple model that reduces the number of 
particles, based on the observed radial distribution in the data, could 
simultaneously improve the model agreement with multiple unfolded 
substructure distributions. Given these observations, the choice was 
made to modify the {\sc pythia} output to produce a reference that 
better matches the unfolded data:

\begin{itemize}

\item Final-state particles are clustered using the 
FastJet~\cite{Cacciari:2012} \antikt algorithm with jet radius $R=0.3$.

\item The ratio of the unfolded data to {\sc pythia}6 in the $r$ 
distributions for each jet \pt bin is used to randomly remove 
constituent particles from the jet. This removal is applied equally to 
charged and neutral particles. The transverse momentum of the removed 
particles is recorded.

\item To avoid changing the overall shape of the jet cross section as a 
function of jet \pt, the momentum of the remaining constituents is 
rescaled to account for the particles that were removed from the jet.

\end{itemize}

Each step in this process required the generation and simulation of 
complete {\sc pythia}6 event samples as described above. It was found 
that the agreement between the integral of the $\xi$, $j_T$ and $r$ 
distributions were in good agreement after two iterations. This event 
sample is referred to as ``modified {\sc pythia}6'' and the final 
results in what follows were generated by unfolding using the modified 
{\sc pythia}6 event sample. The $z_g$ distribution is relatively 
insensitive to the {\sc pythia}6 model used in the unfolding, as 
expected by its construction. Note that although the {\sc pythia} output 
was modified using the radial distribution of particles in the jet, the 
procedure also improves the agreement for the $\xi$ and $j_T/p_{T}^{\rm 
jet}$ distributions as well. The approach chosen to modify {\sc 
pythia}6 by reducing the number of constituents and rescaling their 
momentum to keep the jet momentum unchanged produces a harder 
fragmentation spectrum preferred by the data, as can be seen in 
Fig.~\ref{fig:pythia68}.

\begin{figure}[thb]
\includegraphics[width=0.98\linewidth]{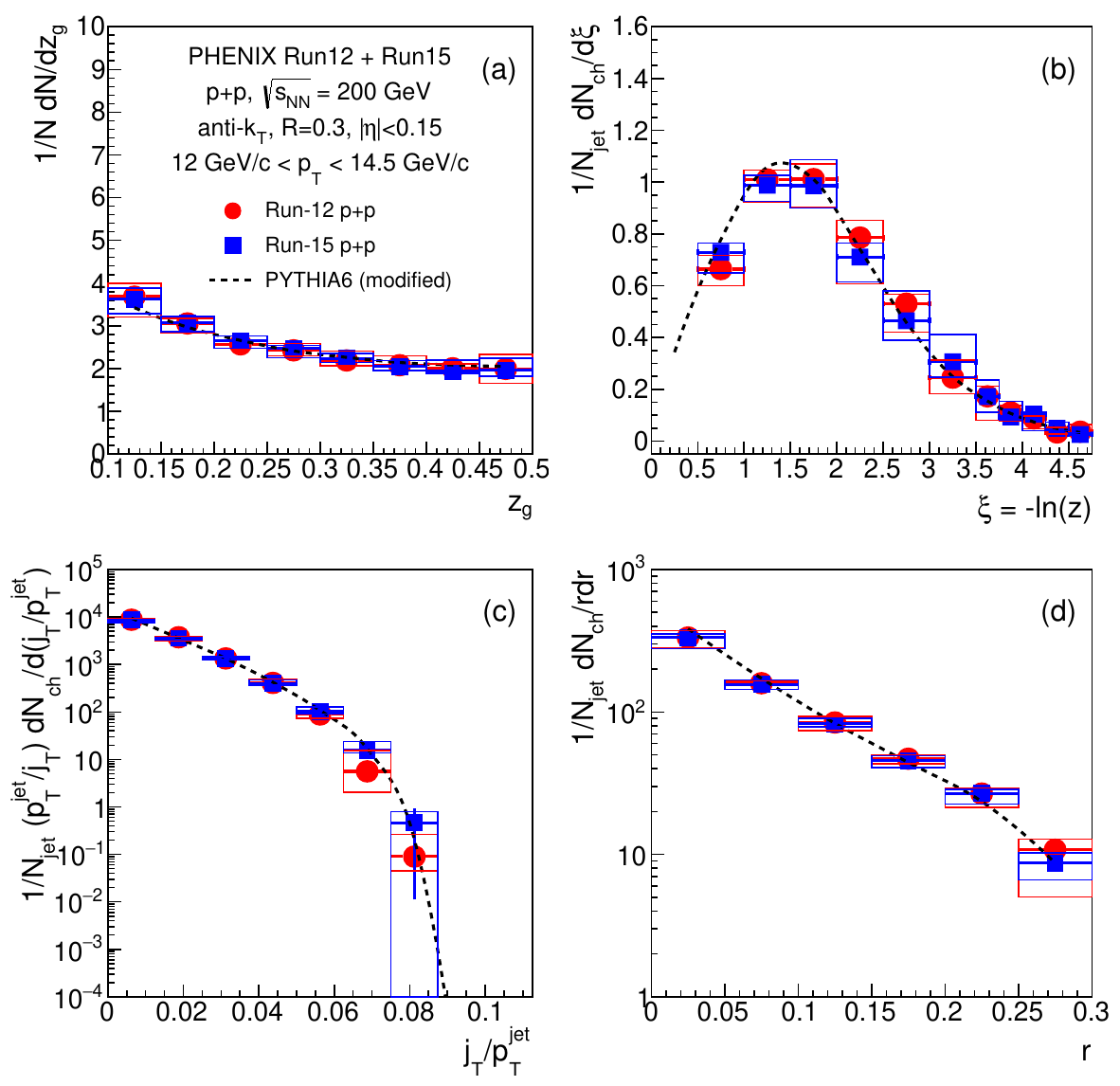}
\caption{\label{fig:comprun12andrun15} 
Comparison of Run12 and Run15 results for four unfolded substructure 
distributions obtained using modified {\sc pythia}6 unfolding matrices. 
The boxes show the systematic uncertainties on the unfolded data points, 
exclusive of the model systematic described in the text.}
\end{figure}

The consistency of the substructure distributions extracted separately 
from the 2012 and 2015 data and unfolded using the modified 
{\sc pythia}6 reference is shown in Fig.~\ref{fig:comprun12andrun15}. 
The final results are produced by combining the 2012 and 2015 
distributions using the full correlation matrix extracted from the 
separate unfoldings to produce the final combined result.

\subsection{Systematic uncertainties}

Systematic uncertainties were calculated for each run period by 
comparing variations of cuts, efficiencies, and unfolding 
procedure to the baseline. These variations included:

\begin{itemize}

\item The Bayesian regulation parameter in the unfolding was varied 
from the nominal 2 to 3 iterations

\item The charged fraction cut on jets was tightened to 0.3--0.6 (from 
0.3--0.7) and the number of constituents ($nc$) cut was raised from $nc 
\geq 3$ to $nc \geq 5$.

\item The outermost pad chamber or EMCal cluster matching cut for drift 
chamber tracks was lowered to 1.5$\sigma$ (from 3$\sigma$). This is our 
dominant source of tracking inefficiency.

\item The minimum \pt of tracks used for jet finding was raised to 1.5 
GeV/$c$, keeping the cluster energy cut at 0.5 GeV/$c$.

\item The minimum energy of clusters used for jet finding was raised to 
1.5 GeV, keeping the track \pt cut at 0.5 GeV/$c$.

\item The energy of EMCal clusters is varied up and down by the scale 
uncertainty of $\pm$3\%, as determined by measurements of $\pi^{0}$ 
mesons

\item The \pt of charged tracks is varied up and down by $\pm$2\%, 
consistent with the estimated track momentum scale uncertainty in 
PHENIX.

\item The overall trigger efficiency was varied within uncertainties.

\item The difference between a separate reconstruction of the east and 
west detector arm results is also added in quadrature to the systematic 
uncertainty, although this difference was negligible.

\end{itemize}

To determine systematic uncertainties, the variations in cuts are 
applied to both the data and the modified {\sc pythia}-MC generator 
processed through the {\sc geant}-based PHENIX simulation. New 
unfolding matrices are generated, the data is unfolded again, and 
the results are compared with the baseline. For energy- and 
momentum-scale errors, the energy scale in the data is shifted and 
the results are unfolded with the standard unfolding matrix. At low 
jet \pt different sources of systematic uncertainty are comparable, 
while at highest \pt the uncertainties related to the unfolding 
procedure dominate. For each run period the systematics determined 
in this fashion are assumed to be uncorrelated and are combined in 
quadrature. An additional overall 10\% systematic uncertainty is 
applied to the cross-section measurement based on the uncertainty 
in the cross section measured by the BBC.

The systematic uncertainties for the combined result is produced by 
combining the 2012 and 2015 systematic uncertainties using the full 
correlation matrix extracted from the separate unfoldings, assuming 
that the systematic uncertainties are correlated through the unfolding 
in the same way as the statistical uncertainties.

Finally, a systematic uncertainty based on the model dependence of the 
unfolding procedure is applied by comparing the results unfolded using 
{\sc pythia}8 and {\sc Herwig} to the results unfolded using modified 
{\sc pythia}6. A point-by-point-modeling systematic uncertainty is 
combined in quadrature with the systematic uncertainties described 
above to produce the final systematic uncertainty on each point in the 
cross section and substructure distribution measurements. The modeling 
systematic uncertainty is subdominant for all but the two highest jet 
$p_T$ points in the cross-section measurements. However, the modeling 
systematic uncertainty dominates for most points in the substructure 
distributions, depending on the specific distribution and jet $p_T$ 
bin.

\section{Results}

\subsection{Jet cross section}

The jet cross section is calculated as
\begin{equation}
    \frac{d^{2}\sigma}{dp_{T} d\eta} = \frac{\sigma_{\rm BBC}}{c^{hard}_{\rm BBC} N_{\rm MB}} \frac{N_{\rm jet}(p_{T})}{\Delta p_{T} \Delta \eta},
\end{equation}
where $\sigma_{\rm BBC}=23.0{\pm}2.2$ mb is the MB cross section 
sampled by the BBC, $c^{\rm hard}_{\rm BBC}=0.79{\pm}0.02$ is the 
correction factor to account for the BBC sampling a larger fraction of 
the cross section when the collision includes a hard scattering 
process. $N_{\rm MB}$ is the effective number of MB events 
sampled by the trigger that pass event-level cuts in offline analysis 
($|z_{\rm vertex}|<10$~cm).

\begin{figure}[thb]
\includegraphics[width=1.0\linewidth]{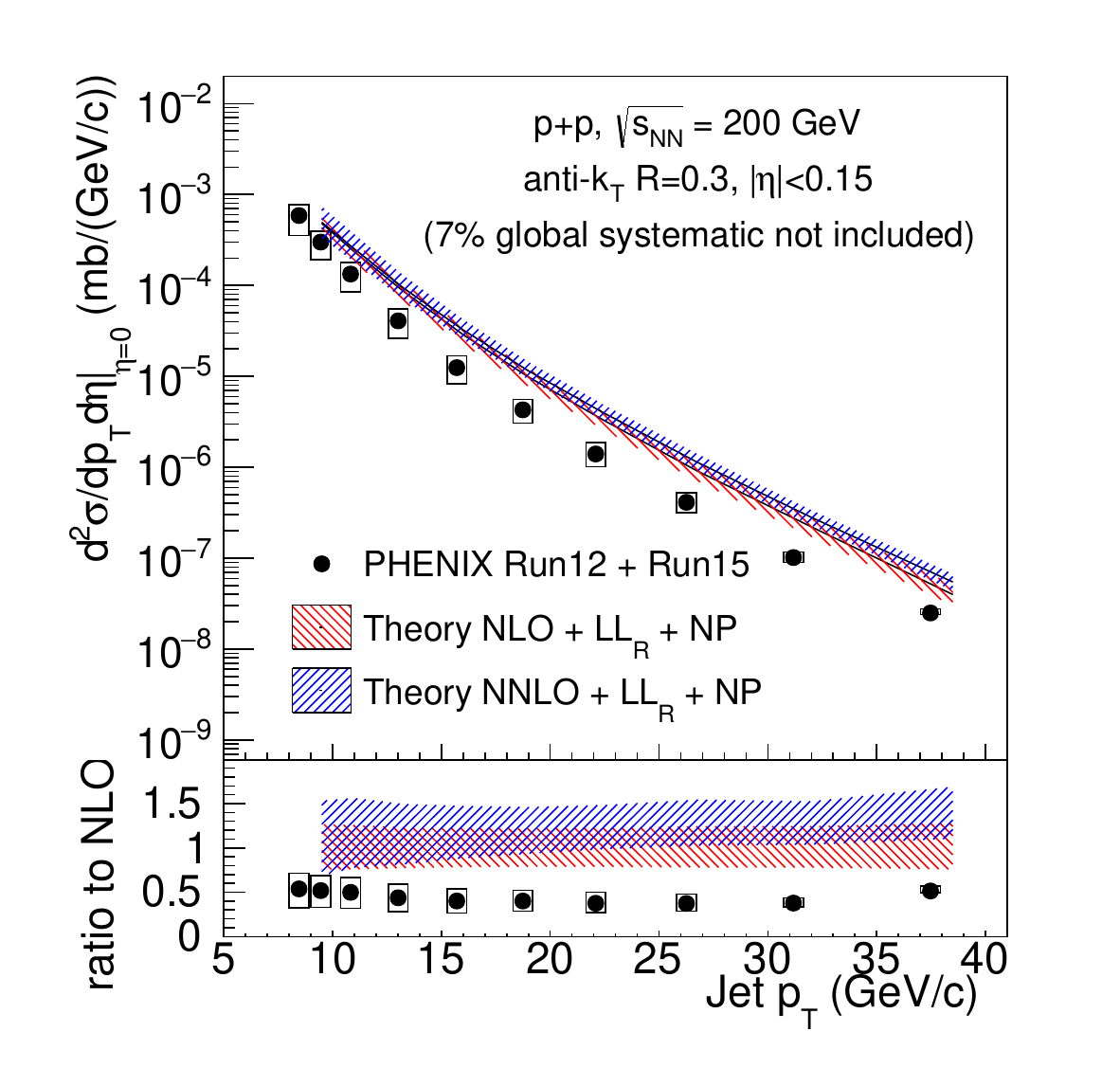}
\caption{\label{fig:cs} 
The jet differential cross section as a function of jet \pt. 
Statistical uncertainties are typically smaller than the data points 
while systematic uncertainties are shown with boxes. An overall 
normalization systematic of 7\% is not included in the point-by-point 
systematic uncertainties. The bottom panel shows the ratio of the data 
and NNLO calculations to the NLO calculations. The theory bands are 
explained in the text and were obtained by matching the NLO and NNLO 
predictions including matching to leading-logarithmic resummation 
($LL_R$) in the jet radius and nonperturbative corrections ($NP$).
}
\end{figure}

The jet differential cross section in \pp collisions at 
$\sqrt{s}=$200~GeV as a
function of \pt is shown in Fig.~\ref{fig:cs}. 
The bands in Fig.~\ref{fig:cs} show theoretical 
calculations obtained by matching the NLO~\cite{Catani1996,Catani1997} 
and NNLO predictions~\cite{Currie2017} to leading-logarithmic 
resummation of the jet radius~\cite{Dasgupta2015}.
The matching is done using the approach 
described in~\cite{Dasgupta2016}, adopting the partonic scalar sum as 
the central scale choice and using the 7-point rule for uncertainties 
(adding the large-angle and small-angle uncertainties in quadrature). 
The perturbative 
calculations are supplemented with nonperturbative ($NP$) corrections 
extracted from Monte-Carlo simulations, as discussed also 
in~\cite{Dasgupta2016}. These corrections are obtained as the average 
and envelope of 5 setups: {\sc pythia}8.306 with tune 
4C~\cite{ATLAS:2012uec}, {\sc pythia}8.306 with tune 
Monash13~\cite{Skands:2014pea}, {\sc pythia}8.306 with tune 
ATLAS14~\cite{ATLAS:2014alx} (with NNPDF2.3~\cite{Ball:2012cx}), 
Sherpa2.2.11~\cite{Gleisberg:2008ta} (default tune), and
 {\sc Herwig}7.2.0~\cite{Bellm:2019zci} (default tune). The nonperturbative 
corrections include hadronization and multi-parton interactions and 
their uncertainties are added in quadrature to the perturbative scale 
uncertainties.

Figure~\ref{fig:cs} shows that theory substantially overestimates the 
measured cross sections. This observation is consistent within 
systematic uncertainties with a previously published comparison between 
jets measured by the STAR Collaboration at RHIC energies using a 
midpoint-cone algorithm and NLO calculations without leading-logarithm 
resummation~\cite{STAR:2006opb}, as well as results from the ALICE 
Collaboration for the low-\pt jet cross section at a higher 
center-of-mass energy ~\cite{Acharya_2019} when compared to MC 
generators.  However, the ALICE results show a jet $p_T$ dependence 
while the PHENIX ratio is flat as a function of jet $p_T$.  Studies of 
the jet cross section relative to NLO predictions at LHC energies 
indicate that NLO predictions overestimate the jet cross section at 
small \antikt $R$, while the agreement is better at larger values of 
$R$~\cite{Khachatryan2017}. This could indicate that the angular 
distribution of particles in the jet is not accurately reproduced by 
NLO calculations. As noted above, NLO calculations work at the partonic 
level, and use a hadronization model to make a comparison to the 
experimental data measured at the hadron level. The hadronization 
correction effectively shifts partonic jet \pt distributions to lower 
hadronic jet \pt. As shown in Ref.~\cite{Dasgupta2016}, the \pt shift 
of the partonic jet due to the hadronization correction is larger at 
jet momenta lower than LHC energies and there is a substantial 
variation between Monte-Carlo models. The hadronization correction 
could also be substantially affected if the fragmentation of the jet is 
substantially different in data than in the Monte-Carlo models, as 
indicated by the unfolding of the PHENIX data.  This could lead to an 
underprediction of the \pt shift by the hadronization and an 
over-prediction of the theory cross section compared to data.

\subsection{Jet substructure distributions}

\begin{figure*}[thb]
\begin{minipage}{0.48\linewidth}
\includegraphics[width=0.99\linewidth]{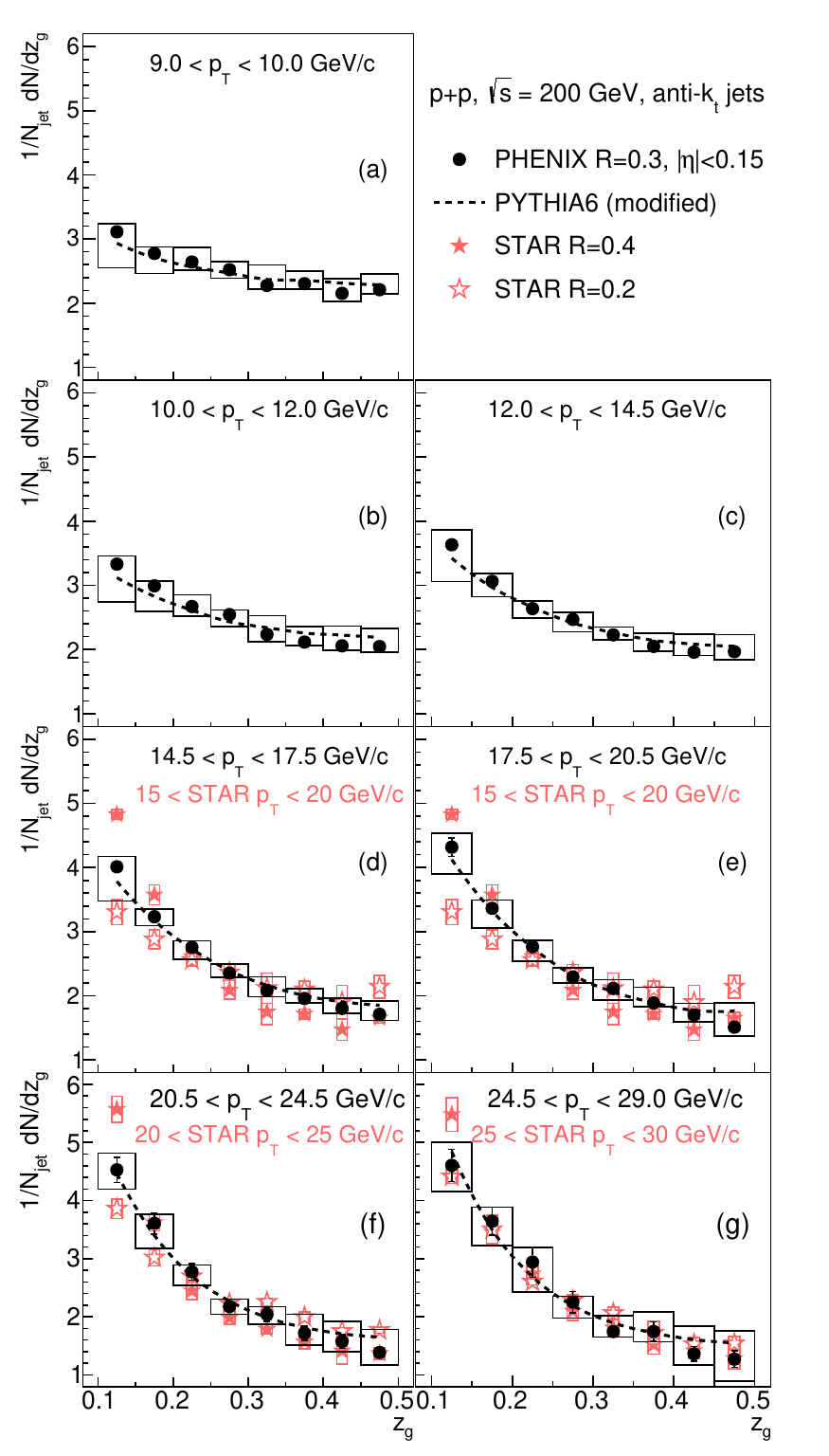}
\caption{\label{fig:zg} 
Distribution of the SoftDrop groomed momentum fraction $z_g$ for 
different jet \pt bins compared to the modified {\sc pythia}6 model 
used in the unfolding and STAR results from~\cite{STARplb}. Standard 
SoftDrop parameters were used ($z_{cut}<0.1$ and $\beta=0$).}
\end{minipage}
\hspace{0.3cm}
\begin{minipage}{0.48\linewidth}
\vspace{-1.05cm}
\includegraphics[width=0.99\linewidth]{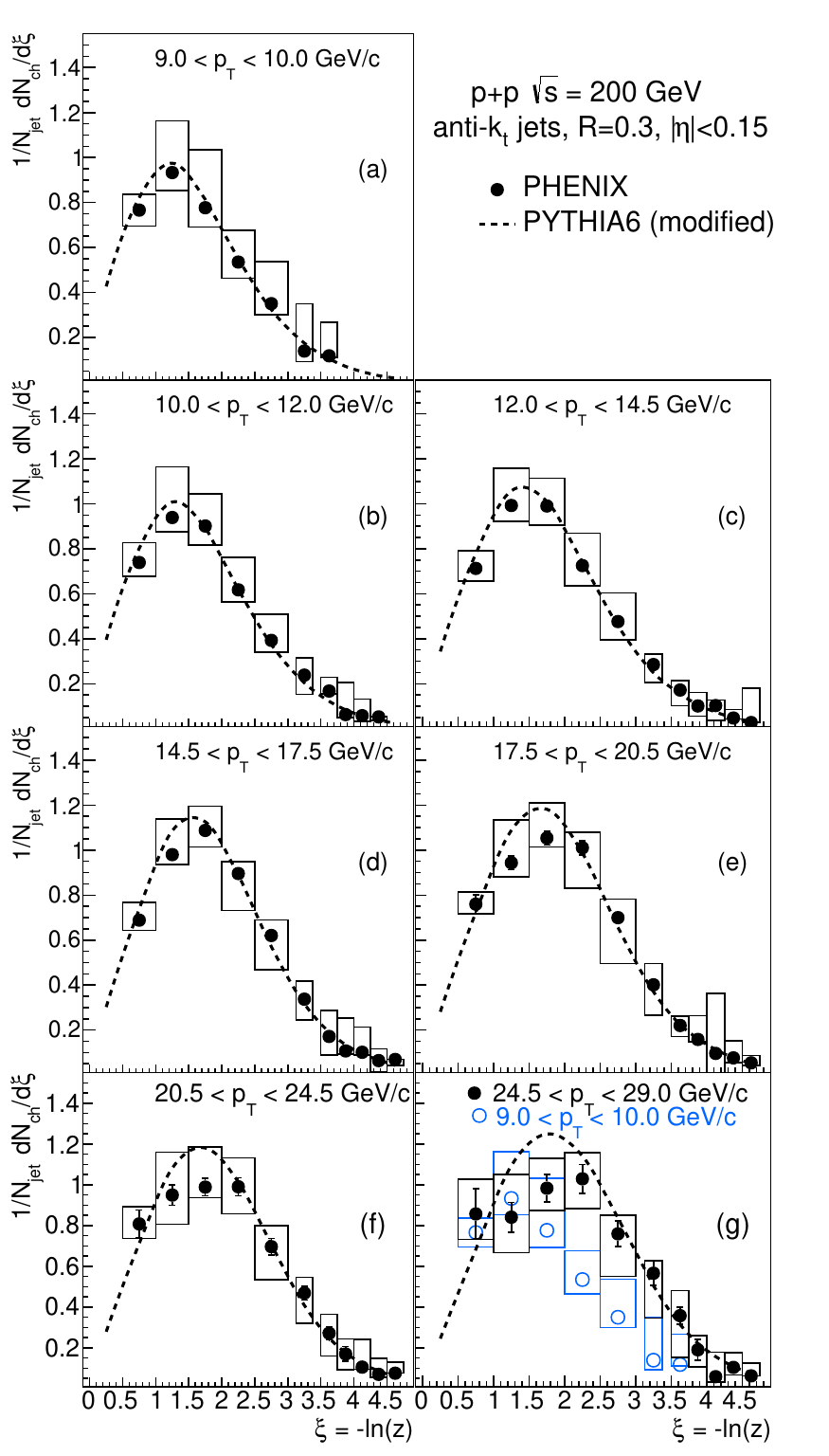}
\caption{\label{fig:xi} 
$\xi$ distributions for different jet \pt bins compared to the modified
{\sc pythia6}~model used in the unfolding.
}
\end{minipage}
\end{figure*}

\begin{figure*}[thb]
\begin{minipage}{0.48\linewidth}
\includegraphics[width=0.99\linewidth]{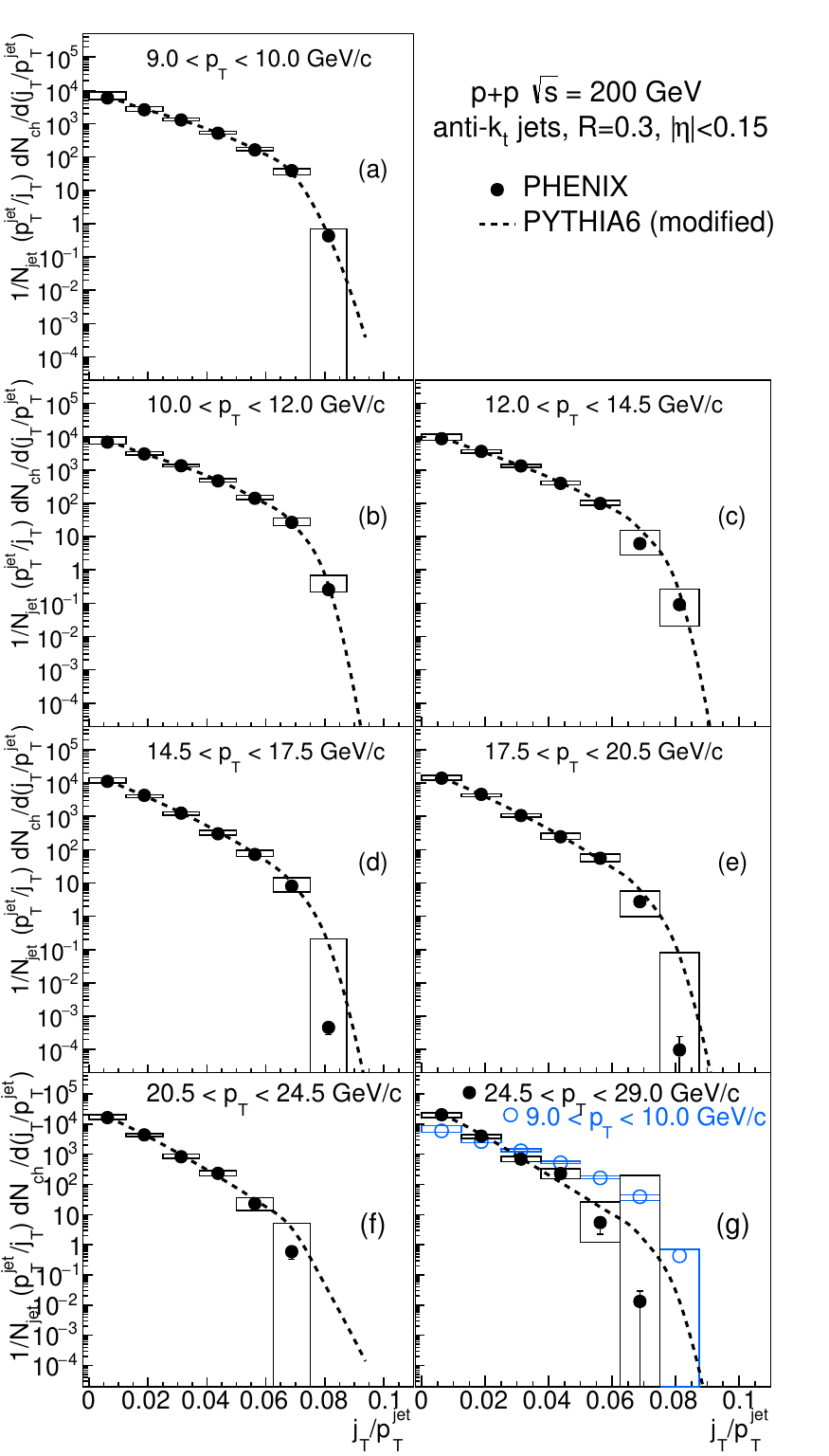}
\caption{\label{fig:jt} 
The $j_T/p_{T}^{\rm jet}$ distributions for different jet \pt bins compared 
to the modified {\sc pythia6}~model used in the unfolding.
}
\end{minipage}
\vspace{-0.2cm}
\hspace{0.3cm}
\begin{minipage}{0.48\linewidth}
\includegraphics[width=0.99\linewidth]{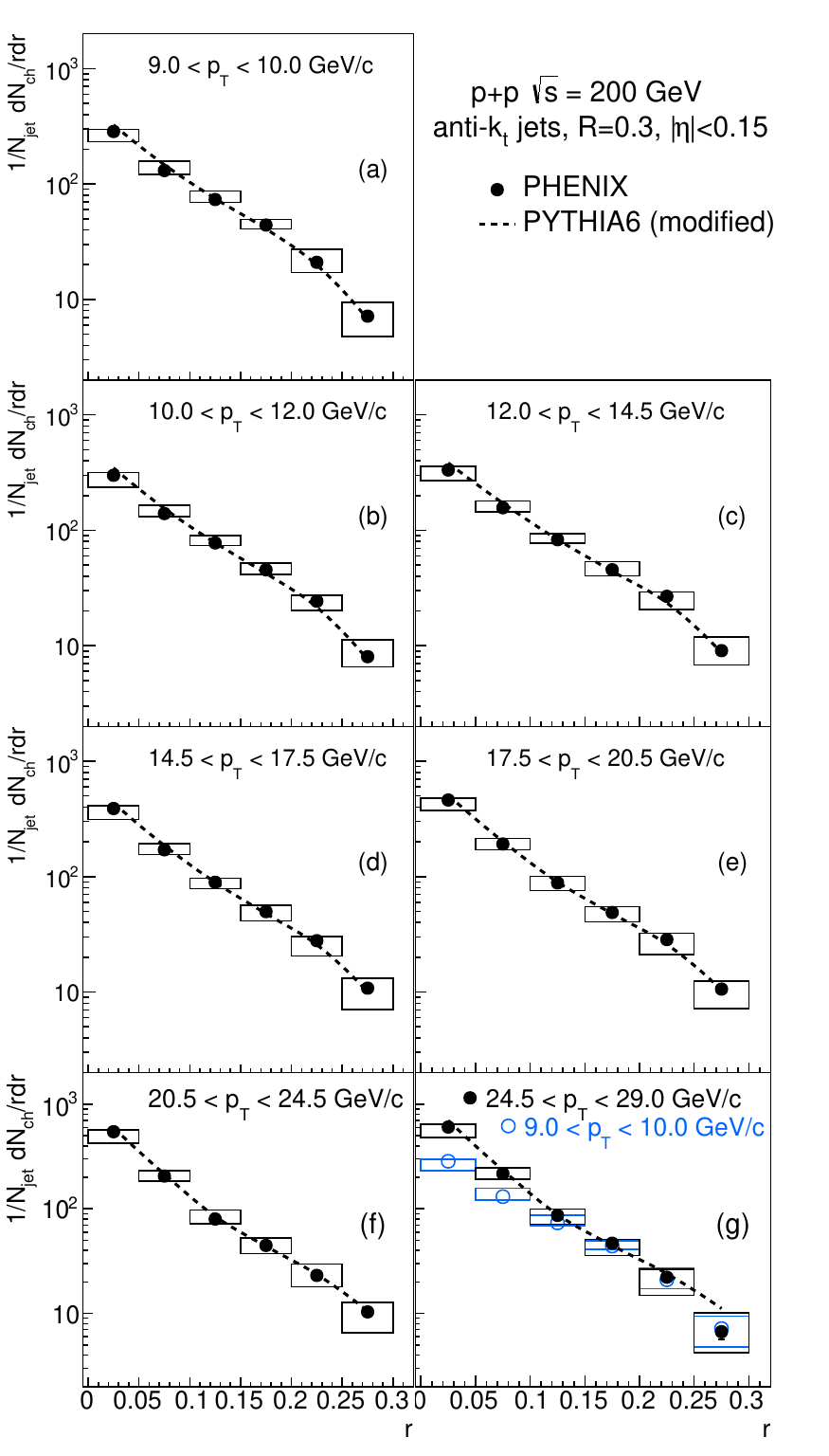}
\caption{\label{fig:dr} 
$r$ distributions for different jet \pt bins compared to the modified 
{\sc pythia6}~model used in the unfolding.
}
\end{minipage}
\end{figure*}

The $z_g$ distribution is calculated using all jet constituents, while 
the distributions in $\xi, j_{T}$ and $r$ are calculated for charged 
particles only.  To derive $z_g$ from a previously determined $R=0.3$ 
\antikt jet, the jet constituents are reclustered using the 
Cambridge-Aachen algorithm~\cite{CAalgo}. This algorithm works by 
clustering from small angles to larger angles, and the clustering tree 
can be accessed to determine the last two sub-clusters that were 
combined to determine the final jet. The quantity $z_g = 
\frac{{\rm min}(p_{T1},p_{T2})}{p_{T1} + p_{T2}}$, where $p_{T1}$ and 
$p_{T2}$ are sub-cluster transverse momenta, is evaluated, and if $z_g 
\leq 0.1$ the lowest \pt cluster is dropped and the remaining subjet 
is declustered 
and evaluated again. This continues until the condition $z_g > 0.1$ is 
met or the jet runs out of constituents. The SoftDrop $z_g$ was first 
measured by the CMS Collaboration in $p$$+$$p$ and Pb$+$Pb collisions at 
$\sqrt{s}=5.02$ TeV at the LHC for jets with $p_T> 140$ 
GeV/$c$~\cite{CMS}, and later by the STAR Collaboration at RHIC 
energies~\cite{STARplb, STARprc}. Figure~\ref{fig:zg} shows 
the SoftDrop~\cite{Dasgupta:2013, 
Larkoski:2014} groomed momentum fraction $z_g$, with SoftDrop condition 
$z_{\rm cut}=0.1$ and SoftDrop $\beta=0$ 
for different \pt bins and the STAR 
results~\cite{STARplb} for different values of \antikt $R$. The STAR 
results are in good qualitative agreement with the PHENIX data. With 
increasing jet \pt the distributions get steeper, demonstrating that jets 
with highly asymmetric splittings are enhanced.

Figure~\ref{fig:xi} shows the distribution of charged particles as a 
function of $\xi = -ln(z)$, where $z$ is the fraction of the jet 
momentum carried by the charged particle, for different \pt bins.  This 
distribution is typically referred to as the fragmentation function. As 
the jet \pt increases, the observed $\xi$ distributions shift right, or 
to smaller constituent momentum fraction $z$.  This is highlighted in 
Fig.~\ref{fig:xi}(g), which compares the lowest and highest jet 
momenta. The PHENIX measurements are limited to $\xi>0.6$ by the 
jet-charged-momentum-fraction cut. A deficit of charged particles in the 
jet relative to the modified {\sc pythia6}~model grows as a function of 
jet \pt between $1<\xi<2.5$.

Figure~\ref{fig:jt} shows the distribution of the charged particle 
transverse momentum with respect to the jet axis $j_T/p_{T}^{\rm jet}$, 
where $p_{T}^{\rm jet}$ is the jet transverse momentum, for different jet 
\pt bins. Figure~\ref{fig:jt}(g), which compares the lowest and highest 
jet-momenta bins, indicates that $j_T$ scales up with increasing jet 
$p_T$ slower than the jet $p_T$ itself.  This is consistent with the 
changes observed in the $r$ distribution.

The radial distribution of charged particles within the jet with 
respect to the jet axis ($r$) is shown in Fig.~\ref{fig:dr} as a 
function of jet \pt. The distribution of particles in the jet as a 
function of distance from the jet axis shows a significant increase at 
small $r$ with increasing jet \pt, as can be seen in 
Fig.~\ref{fig:dr}(g), where both the lowest and highest jet \pt bins 
are superimposed.  This indicates the development of a higher particle 
density in the core of the jet with increasing jet \pt, which is 
consistent with the expected increase in the contribution of quark jets over 
gluon jets with increasing jet \pt at RHIC~\cite{Adam2019}.

\section{Summary and Conclusions}

In summary, presented here are the jet $p_T$-differential cross section 
and jet substructure distributions in \pp collisions at \sqs=200 GeV. 
Jets were reconstructed using the \antikt algorithm with a jet radius 
$R=0.3$ for jets with transverse momentum within $8.0<p_T<40.0$ GeV/$c$ 
and pseudorapidity $|\eta|<0.15$. The results were unfolded for 
experimental and detector effects.  The unfolding indicates a lower 
average charged particle multiplicity is observed in the PHENIX data 
than in the {\sc pythia} event generators, as much as one particle at 
the highest measured jet \pt.

These results indicate that NLO and NNLO predictions are higher than the measured jet cross section at RHIC, a result that is within the large systematic errors in a prior measurement~\cite{STAR:2006opb}. This may indicate a limitation of the procedure used to translate 
from the partonic to the hadronic cross section, which requires 
Monte-Carlo generators for the nonperturbative (NP) corrections. The 
measured data indicates a lower particle multiplicity at these 
center-of-mass energies and jet momenta than in the event generators 
used to calculate these corrections, while measurements at the LHC 
indicate that NLO calculations overestimate the jet cross section at 
small \antikt $R$. This indicates there may be multiple effects 
contributing to the disagreement between QCD calculations of the jet 
cross section and the measured data.

Presented were unfolded distributions in jets for $z_g, \xi, 
j_T/p_{T}^{\rm jet},$ and $r$. The measured $z_g$ distribution agrees well 
with the STAR results and becomes steeper with increasing jet \pt.  The 
$\xi$ distribution shifts towards lower momentum fraction within the 
range measured in the PHENIX data. The $j_T/p_{T}^{\rm jet}$ distribution 
stays relatively unchanged with increasing jet \pt, while the $r$ 
distribution shows a significant increase at small $r$ with increasing 
jet \pt, consistent with an increasing fraction of quark jets at higher 
jet \pt.

In conclusion, these measurements contribute to an improved 
understanding of the jet cross section and substructure in \pp 
collisions at RHIC, and are essential to be able to exploit new data 
from the sPHENIX detector, which will measure jets in heavy-ion 
collisions at RHIC with unprecedented precision~\cite{sPhenix2022}. In 
addition, as the center-of-mass energies and \pt range will be similar 
these results will also help inform jet measurements at the future 
Electron-Ion Collider.






\begin{acknowledgments}

We thank the staff of the Collider-Accelerator and Physics
Departments at Brookhaven National Laboratory and the staff of
the other PHENIX participating institutions for their vital
contributions.  
We acknowledge support from the Office of Nuclear Physics in the
Office of Science of the Department of Energy,
the National Science Foundation,
Abilene Christian University Research Council,
Research Foundation of SUNY, and
Dean of the College of Arts and Sciences, Vanderbilt University
(USA),
Ministry of Education, Culture, Sports, Science, and Technology
and the Japan Society for the Promotion of Science (Japan),
Natural Science Foundation of China (People's Republic of China),
Croatian Science Foundation and
Ministry of Science and Education (Croatia),
Ministry of Education, Youth and Sports (Czech Republic),
Centre National de la Recherche Scientifique, Commissariat
{\`a} l'{\'E}nergie Atomique, and Institut National de Physique
Nucl{\'e}aire et de Physique des Particules (France),
J. Bolyai Research Scholarship, EFOP, the New National Excellence
Program ({\'U}NKP), NKFIH, and OTKA (Hungary),
Department of Atomic Energy and Department of Science and Technology
(India),
Israel Science Foundation (Israel),
Basic Science Research and SRC(CENuM) Programs through NRF
funded by the Ministry of Education and the Ministry of
Science and ICT (Korea),
Ministry of Education and Science, Russian Academy of Sciences,
Federal Agency of Atomic Energy (Russia),
VR and Wallenberg Foundation (Sweden),
University of Zambia, the Government of the Republic of Zambia (Zambia),
the U.S. Civilian Research and Development Foundation for the
Independent States of the Former Soviet Union,
the Hungarian American Enterprise Scholarship Fund,
the US-Hungarian Fulbright Foundation,
and the US-Israel Binational Science Foundation.

\end{acknowledgments}

\section*{DATA AVAILABILITY}

The data that support the findings of this article are not publicly available.
The values in the plots and tables associated with this article are stored in
HEPData~\cite{hepdata}.


%
 
\end{document}